# Ultrafast pulse generation with black phosphorus

*Diao Li[1,2][‡], Henri Jussila[1][‡], Lasse Karvonen[1], Guojun Ye[3,4], Harri Lipsanen[1],*

*Xianhui Chen[3,4,5], and Zhipei Sun[1][*]*

[1] Department of Micro- and Nanosciences, Aalto University, Tietotie 3, FI-02150 Espoo, Finland

[2] Institute of Photonics & Photo-Technology, Northwest University, Xi'an, 710069, China

[3] Hefei National Laboratory for Physical Science at Microscale and Department of Physics, University of Science and Technology of China, Hefei, Anhui 230026, China

[4] Key Laboratory of Strongly-coupled Quantum Matter Physics, University of Science and Technology of China, Chinese Academy of Sciences, Hefei, Anhui 230026, China

[5] Collaborative Innovation Center of Advanced Microstructures, Nanjing 210093, China

**ABSTRACT:** Black phosphorus has been recently rediscovered as a new and interesting two-dimensional material due to its unique electronic and optical properties. Here, we study the linear and nonlinear optical properties of black phosphorus thin films, indicating that both linear and nonlinear optical properties are anisotropic and can be tuned by the film thickness. Then we employ the nonlinear optical property of black phosphorus for ultrafast (pulse duration down to ~786 fs in mode-locking) and large-energy (pulse energy up to >18 nJ in Q-switching) pulse generation in fiber lasers at the near-infrared telecommunication band ~1.5 μm. Our results underscore relatively large optical nonlinearity in BP and its prospective for ultrafast pulse generation, paving the way to BP based nonlinear and ultrafast photonics applications (e.g., ultrafast all-optical switches/modulators, frequency converters etc.).

**INTRODUCTION**

Pulsed laser sources are used in a variety of applications [1-3], ranging from basic research to telecommunications, medicine, and industrial materials processing [1-3]. The most-widely used pulsed lasers utilize a Q-switching method or a mode-locking technique [1-3], in which a typical nonlinear optical device, called saturable absorber (SA), turns the continuous wave output of the laser into a periodic train of optical pulses. The SA technology is currently dominated by semiconductor saturable absorber mirrors (SESAMs) [1-3]. However, they typically have limited bandwidth and require complex fabrication and packaging [1]. Recently, carbon nanotubes (CNTs) [4] and graphene [5-6] have been demonstrated for SAs with superior performances [7-10], such as broad operation bandwidth [11-13], fast recovery times [14-17], low saturation intensity [4-17], cost-effective and easy fabrication [4-17]. Nevertheless, SAs based on these materials still suffer from drawbacks. For example: when operating at a particular wavelength, CNTs which are not in resonance cannot be used, and thereby give relatively large insertion losses [5-7,18]; On the other hand, mono-layer graphene typically has rather weak absorption (~2.3% [19-20]), not suitable for various lasers (e.g., fiber lasers), which typically need relatively larger modulation depth [7-9]; Layered transition metal dichalcogenides (TMDs) (e.g., $MoS_2$,[21] $WS_2$,[22] and $MoSe_2$ [23-24]) have also been demonstrated for SAs, but with limited performance for current lasers typically operating at the near-infrared and mid-infrared range, due to their comparatively large bandgap near or in the visible region [25] (~1.8 eV for $MoS_2$, ~2.1 eV for $WS_2$, ~1.7 eV for $WSe_2$ [26]).

Black phosphorus (BP), a layered material consisting of only phosphorus atoms, has recently been rediscovered for various electronics and optoelectronics applications [27-52] (such as transistors, solar cells, and photodetectors). In contrast to graphene and TMDs, BP has its own unique properties [27-52]. For example, its direct electronic band gap can be tuned from ~0.3 eV to

~2 eV (corresponding to the wavelength range from ~4 to ~0.6 μm), depending on the film thickness [27-52]. This is particularly interesting for photonics, as it can offer a broad tuning range of band gap with number of layers for the near and mid-infrared photonics and optoelectronics, and bridge the present gap between the zero band gap of graphene and the relatively large band gap of the most studied TMDs [32].

However, thus far, intensive research efforts published on BP have mainly focused on its electronic properties (e.g., transistor performance) and linear optical response (e.g., photo-detector performance). In this paper, we investigate the thickness and polarization dependent linear and nonlinear optical properties of BP thin films, which are integrated into fiber devices, the most commonly-used format for optical telecommunications. Our results show that both linear and nonlinear absorption properties are strongly thickness/polarization dependent. We also demonstrate the use of nonlinear optical property of BP for ultrafast (pulse duration down to ~786 fs in mode-locking) and large-energy (pulse energy up to >18 nJ in Q-switching) pulse generation in fiber lasers at the near-infrared telecommunication band ~1.55 μm. Our results pave the way to black phosphorus based nonlinear and ultrafast photonics applications (e.g., ultrafast optical switches/modulators, frequency converters etc).

**RESULTS AND DISCUSSION**

BP thin films are produced by micromechanical cleavage of a bulk BP crystal, and transferred to optical fiber ends (details in Methods). The thicknesses of the transferred films on fiber ends were then measured by Atomic Force Microscopy (AFM). Figures 1(a) and 1(b) show AFM image taken from a typical BP film and its line profile along the dashed white line. The circular fiber cladding can be resolved from the image, and the location corresponding to the fiber core (marked with the green circle) of a standard single mode fiber (Corning SMF-28, with a core diameter of ~10 microns) is drawn schematically in Fig. 1(a). The thickness of the transferred BP film is estimated to be ~25 nm at the location corresponding to the fiber core (Fig. 1(b)). Typically, the thickness of transferred BP films ranges between ~20 nm and ~1 μm, depending on the micromechanical cleavage process. To verify that the transferred material is BP, we perform polarization-resolved Raman scattering measurements. Raman spectrum of a BP crystal is depicted in Fig. 1(c). Three peaks located at the wavenumber of 363 cm$^{-1}$, 441 cm$^{-1}$ and 469 cm$^{-1}$ can be observed from the Raman spectrum, and attributed to $A_g^1$, $B_{2g}$ and $A_g^2$ vibration modes of BP crystal lattice, respectively. This agrees well with previously published results on BP thin-films [29,30,53]. The Raman peak intensity is also strongly dependent on excitation light polarization (Supporting Information Fig. S2) due to its highly anisotropic optical responses [30], and this has been noted to offer a unique method for determining the crystal orientation of BP films [30,44].

We then characterize the linear and nonlinear absorption properties of BP films transferred to the optical fiber ends. The linear absorption results acquired at 642 nm (~1.93 eV, Fig.2 (a)) and 520 nm (~2.38 eV, Fig.2 (b)) show that the transmittance of BP thin films decreases with the increase of the film thickness. As shown in Figs. 2(a) and 2(b), the transmittance (T) agrees well with the fit (the solid lines) using the Beer-Lambert law (i.e., $T=\sim\exp(-\alpha\times d)$ where α is the absorption coefficient and d is the film thickness) with the fitted values of $\alpha_{642nm}$= ~5.7 μm$^{-1}$, $\alpha_{520nm}$= ~10 μm$^{-1}$. Thanks to the availability of our polarization-tuneable continuous-wave light source at 1.55 μm (~0.8 eV), we measure the transmittance change of our BP films as a function of incident light polarization angle at this wavelength (i.e. 1.55 μm). The results from 25 nm and 1100 nm thick BP films are given in Fig 2 (c). It appears that the polarization state affects strongly to the relative magnitude of BP absorption (and thus the transmittance). For instance, we observe that the transmittance of the 1100 nm thick BP film can increase by a factor of >9 (from 3.6% to 33.3%) when the polarization state is altered. It clearly shows the absorption anisotropy of BP [35-36,49,52]. The polarization directions corresponding to the maximum and minimum transmittance are along the zigzag and armchair directions of BP thin films [35-36], respectively. Therefore, such absorption anisotropy property can be employed to determine the crystal orientation of BP films, similarly to the Raman approach [30,44]. Worth noting that this property can be utilized directly for various polarization-based photonic applications (e.g., polarizers).

We find that the polarization dependent transmittance change is significantly larger in thicker samples. For example, the transmittance change (~29.6%) of the 1100 nm thick sample is >6 times larger than the result (~4.8%) of the 25 nm sample (Fig. 2 (c)). Detailed transmittance of samples with variable thicknesses at two orthogonal polarized light directions (Fig. 2 (d)) further confirms that the polarization-introduced transmittance change is thickness dependent, which is linked to the selection rules associated with symmetries of the anisotropic material [35-36,52]. At this wavelength (i.e., 1.55 μm, Fig. 2 (d)), we also observe that the film thickness dependent transmittance matches well with a bi-exponential decay fit which contains two different absorption coefficients, in contrast to the single exponential decay fit of using the Beer-Lambert law at the wavelengths of 642 nm and 520 nm (Figs. 2(a) and 2(b)). As depicted in Fig. 2 (d), the transmittance decreases first rapidly until the thickness is ~80 nm. After that, the transmittance decreases slowly. We assume that such a large difference in absorption coefficient is most likely due to the band gap tuning effect of BP with the increasing film thickness. The change in the band gap ($E_g$) of BP has been predicted to follow a power law (e.g., $E_g \approx \frac{1.7\text{eV}}{n^{0.73}} + 0.3\text{eV}$, in which n is the number of layers) [36,47]. Therefore, the change in bandgap attributable to the increasing film thickness can be deduced to be extremely small (compared to the 0.8 eV photon energy used in this experiment), when the sample is very thick (~>80 nm). Consequently, we believe the small absorption coefficient for thick samples (>~80 nm) follows the Beer-Lambert law, similarly to what we observed for the relatively large photon energy transmittance measurement experiments (642 nm in Fig. 2(a), and 520 nm in Fig. 2(b)).

The nonlinear absorption measurement results are illustrated in Figs. 2(e) and 2(f). In our measurement setup (Supporting Information Fig. S4), we placed a polarization controller before the BP films to adjust the polarization direction of the input ultrafast pulses. Figure 2(e) depicts the nonlinear absorption measurement results of an 1100-nm thick BP film with two orthogonal polarization directions. A clear increase in the transmittance with the increased pump fluence can be observed in the 1100 nm thick BP sample and is attributed to saturable absorption [42]. The polarization dependent nonlinear optical performance difference is also observed in Fig. 2(e), which will be interesting for various photonic applications [54], e.g., tuning operation states in ultrafast lasers or switching optical pulses with their polarization directions. Figure 2(f) shows the relative transmittance increase ($\Delta T/T_0$, where $\Delta T$ and $T_0$ are transmittance change and the transmittance at the minimum input power, respectively) for three BP films with the polarization state corresponding to the maximum absorbance (i.e., the armchair-polarized input). Nonlinear saturable absorption is clearly observed in all samples and occurs when the fluence reaches to ~100 µJ/cm$^2$. We also note that the thicker sample has ~8-time larger relative transmittance increase than the thinner one. This shows that the nonlinear property of BP can be adjusted by the thickness (i.e., number of layers). To estimate the saturation fluence and modulation depth from the nonlinear absorption curves, we use a simplified fluence dependent absorption formula to fit the measurement results (descripted in Supporting Information). The fitted curves match decently with the measurement results and are plotted with solid lines in Fig. 2(e) and 2(f). The obtained saturation fluence from all the samples varies in the range of 2000 µJ/cm$^2$ and is, therefore, around an order of magnitude larger than that typically measured with SAs fabricated from CNTs or graphene [6-10,15]. On the other hand, the modulation depth obtained from the measured curves are observed to be larger than 1%. However, the modulation depth obtained from the fits typically ranges between 50% and 90%. If true, this observation is promising as the fitted modulation depths are extremely large. However, we note that the fitted modulation depths are probably unrealistically high and most likely relate to the fact that the nonlinear absorption measurement should be continued to larger fluence range which are currently unavailable in our setup. In our nonlinear absorption measurement setup (Supporting Information Fig S4), the available maximum fluence is ~450 µJ/cm$^2$.

We then use our BP integrated fiber device to build a pulsed fiber laser working at the main telecommunication window of 1.55 µm. Fiber laser is selected in our experiments, as it has a simple and compact design, efficient heat dissipation, and high-quality pulse generation [55,56]. The layout of our designed fiber laser is schematized in Fig. 3(a). A ~1-m erbium-doped fiber (EDF) is utilized as the gain medium, which is pumped by a 980 nm laser diode (LD) via a wavelength division multiplexer (WDM). A polarization-independent isolator (ISO) is placed after the gain fiber to maintain unidirectional operation. A polarization controller (PC) optimizes pulse operation state. A 10/90 coupler is used to extract the light from the cavity for measurements. The total cavity length is ~11 m.

We get Q-switched optical output from the fiber laser, only after inserting the BP integrated device inside the cavity. Q-switching operation is achieved with all BP samples, but the 1100 nm thick BP film gives better performance, as expected from the relatively large transmittance change performance in the device (Fig. 2 (f)). The output performance using the 1100 nm thick

BP film is listed in Fig. 3 (b-d). The threshold pump power for continuous wave lasing is ~11 mW (The output power as a function of pump power is given in Supporting Information Fig. S5). When the pump power is increased to ~23 mW, stable Q-switching can be achieved. The peak wavelength is ~1532.5 nm, with the full width at half maximum (FWHM) of ~3 nm. The output repetition rate and pulse duration is pump-power dependent (Fig. 3 (c)), a typical signature of Q-switching. This is because: when the pump power increases, larger gain is provided to saturate the SA, and thus the repetition rate increases and consequently the pulse duration reduces. In our experiment, the repetition rate increases from ~26 to ~40 kHz, and the pulse duration decreases from ~9.5 to ~3.16 µs, when the pump power is raised from ~23 to ~55 mW. Figure 3 (d) plots a typical pulse train, showing a FWHM pulse-duration of ~3.16 µs, and a pulse period of ~24.8 µs (the corresponding pulse repetition rate is ~40 kHz). The maximum output power in our experiment is 728 µW, with output pulse energy of 18.2 nJ. Note that this output performance is very comparable to typical Erbium-doped fiber lasers Q-switched with other nanomaterials (e.g., CNTs and graphene [7-10]).

When the fiber cavity length is increased to ~14.2 meter (after adding ~3 m of SMF-28 single mode fiber in the laser cavity), the total group velocity dispersion of our fiber cavity is ~ -2.5×10$^{-1}$ ps$^2$. In this case, it can facilitate soliton-like pulse shaping through the interplay of group velocity dispersion and self-phase modulation [56]. Indeed, after inserting our BP integrated fiber device in this fiber cavity, stable mode-locking can be initiated by introducing a disturbance to the intra-cavity fiber. Once stable output is achieved, no further polarization controller adjustment is required. The output power is ~1 mW. Figure 4 summarizes the mode-locked laser performance. The laser mode-locks at 1558.7 nm, with the FWHM of 6.2 nm. The side bands (1546.76, 1551.16, 1566.36, 1570.76, and 1574.36 nm, shown in Fig. 4 (a)) fully confirm our

soliton-like mode-locking, as they are typical of soliton-like pulse formation, resulting from intra-cavity periodical perturbations of discrete loss, gain and dispersion [57]. Figure 4 (b) gives a typical output autocorrelation trace, which is well fitted by a sech$^2$ temporal profile. The pulse duration is ~786 fs. The time-bandwidth product (TBP) of the mode-locked pulses is ~0.6. The deviation from the TBP value of ~0.315 anticipated for transform-limited sech$^2$ pulses suggests the presence of chirping of the generated ultrafast pulses [56]. Shorter pulses may be obtained using fiber lasers with specifically-design dispersion map (e.g., stretched-pulse fiber laser design [14,16,58]).

Figure 4 (c) depicts the output pulse train, with a period of 68.16 ns, corresponding to the cavity fundamental repetition rate $f_0$ of 14.7 MHz, as expected from the total fiber cavity length of ~14.2 meters. To investigate the laser output stability [59-60], we characterize the radio frequency spectrum. We first measure broad-span frequency spectrum up to 500 MHz (Supporting Information Fig. S6). It presents no significant spectral modulation, implying no Q-switching instabilities [59-60]. Figure 4 (d) gives the radio frequency spectrum around the fundamental repetition rate ($f_0$). A >50 dB signal-to-background ratio (corresponding to >$10^5$ contrast) is observed, showing good mode-locking stability [59-60]. Note that the performance of the BP mode-locked laser is comparable to what was typically achieved with CNTs and graphene based fiber lasers [5-10]. However, given the unique bandgap tuning property at the mid-infrared range, we expect superior performance of BP thin films for ultrafast lasers at this spectral range, worthy of future research.

**CONCLUSIONS**

In summary, we have studied the thickness and polarization dependent linear and nonlinear optical properties of BP thin films, and then utilized their nonlinear absorption property to generate ultrafast and large-energy pulse generation with BP integrated fiber devices. Our results exhibit the practical potential of this promising material for various nonlinear and ultrafast optoelectronics applications (e.g., ultrafast lasers, switches and modulators).

**Methods**

*BP device fabrication*: Black phosphorus was synthesized under a constant pressure of 10 kbar by heating red phosphorus to 1,000 °C and slowly cooling to 600 °C at a cooling rate of 100 °C per hour. Red phosphorus was purchased from Aladdin Industrial Corporation with 99.999% metals basis. The high-pressure environment was provided by a cubic-anvil-type apparatus (Riken CAP-07). After that, BP films were produced by micromechanical cleavage of bulk BP crystals directly onto a viscoelastic PDMS stamp. A selected BP film on the PDMS stamp is then placed on a fiber end with the help of optical microscope and micromanipulator. Due to viscoelastic properties of PDMS, the BP film adheres to the fiber end when the PDMS stamp is gently lifted off [45-46].

*BP device characterization*:

AFM *characterization*: AFM measurements were performed in semi-contact mode using NTegra Aura AFM apparatus equipped with a scanning head. A custom-made measurement stage was fabricated allowing us to characterize the BP films attached on the fiber end. The maximum scan size of the setup was $100 \times 100$ μm$^2$.

*Raman spectroscopy:* Raman spectra were performed by using a confocal Raman microscope (Witec alpha 300R) equipped with a frequency doubled Nd:YAG green laser ($\lambda = 532$ nm). The samples are placed on the $SiO_2$/Si substrate, fabricated with the same fabrication approach discussed above, and the thicknesses of the characterized films were measured by AFM.

*Linear absorption measurement:* A home-made erbium-doped fiber based amplified spontaneous emission source was used to characterize the absorption spectrum from ~1500 to 1600 nm. Its output polarization is changed with a prism based polarizer to measure the polarization dependent transmittance. Absorption properties at different wavelengths (e.g., 520, 642 nm) were measured with various fiber coupled non-polarized laser diodes (i.e., without polarization-tuning capability). The input power for the linear absorption measurement was set less than 1mW.

*Nonlinear absorption measurement:* A power-amplified home-made ultrafast fiber laser (~15 mW, 530 fs, 62 MHz) was employed to measure the saturable absorption property of the BP based fiber devices. A polarization controller was used to change the light polarization direction to measure polarization dependent saturable absorption performance. A double channel power meter (Ophir, Laserstar) was used to achieve high-accuracy measurement.

*BP based pulsed laser characterization*: An optical spectrum analyser (Anritsu, MS9740A), a power meter (Ophir, Nova II), and a second-harmonic generation autocorrelator (APE, Pulse-check50) were used to characterize the generated ultrafast pulse performance. Pulse train was measured by an oscilloscope connected with a photodetector, while the radio frequency spectrum was taken by a radio frequency analyser (Anritsu MS2692A) with an ultrafast (>25 GHz) photodetector.

**Figure 1.** (a) AFM image of transferred black phosphorus film on the fiber end. (b) Line profile along the dashed white line (marked in (a)). The thickness of BP film is ~25 nm at the fiber core (marked with a green circle in (a)). (c) Raman spectrum of a typical BP film.

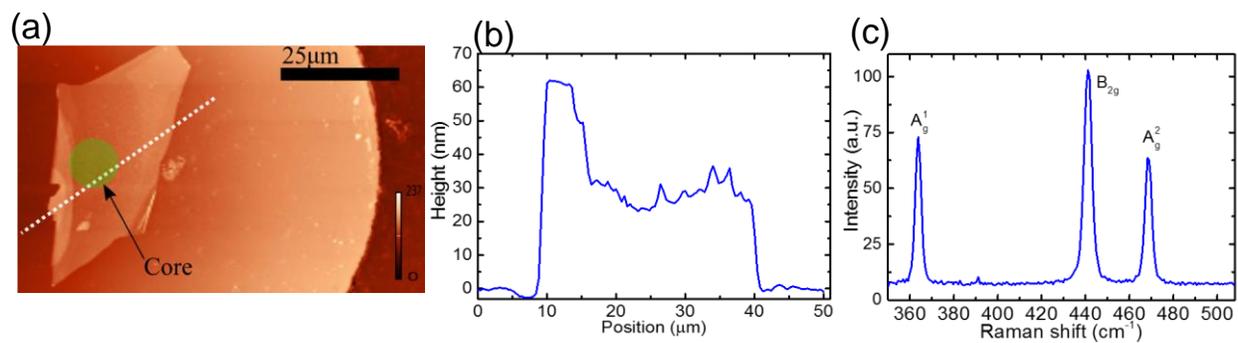

**Figure 2.** Linear and nonlinear optical property of BP films: Transmittance of BP films as a function of thickness at the wavelength of 642 nm (a) and 520 nm (b). (c) Polarization dependent transmittance for 25 nm and 1100 nm thick BP films. The polarization directions corresponding to the maximum and minimum transmittance are linked with the zigzag and armchair axes of BP thin films. (d) Transmittance of BP films as a function of thickness at the wavelength of 1550 nm with two orthogonal polarized light directions; (e) Fluence dependent transmittance of the 1100 nm thick BP film measured with ultrafast pulses at two orthogonal polarized light directions. (f) Relative transmittance increase measured from 25 nm, 350 nm and 1100 nm thick BP films as a function of input pulse fluence. The input polarization direction is along the armchair direction of the BP films.

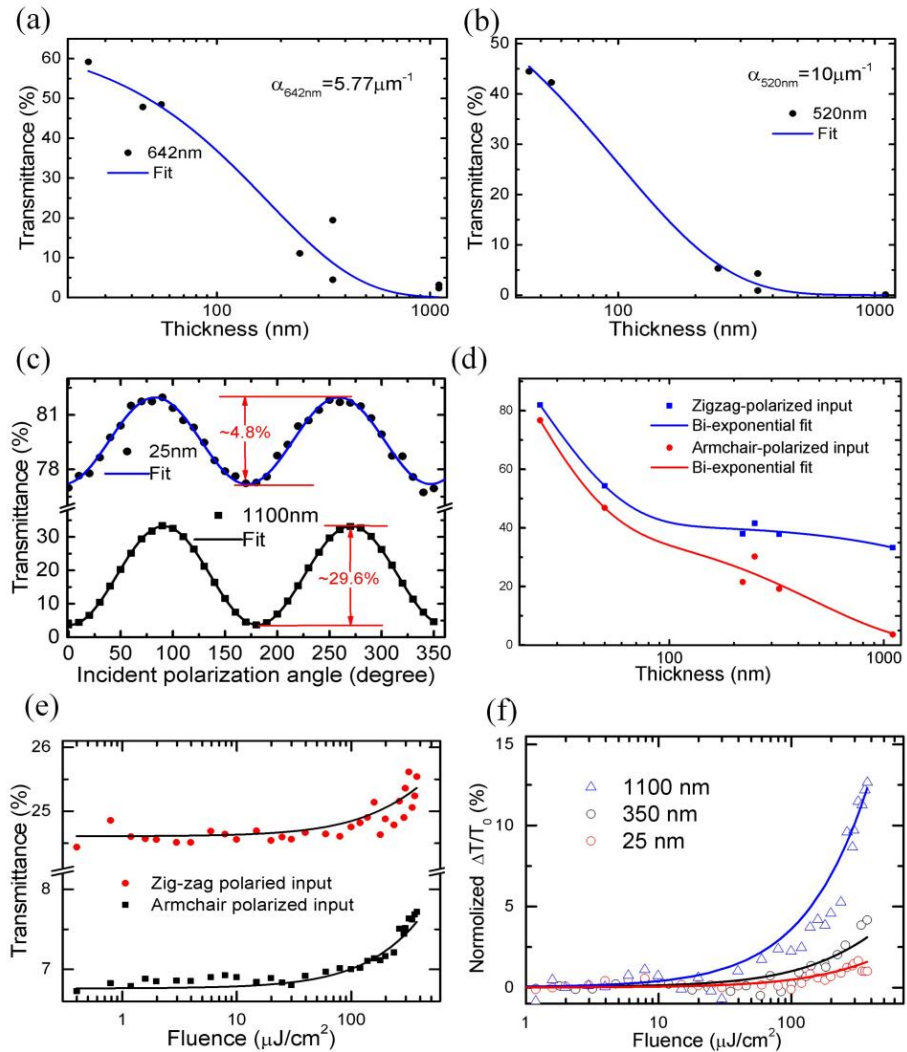

**Figure 3.** BP Q-switched fiber laser: (a) Schematic of the laser setup. PC: polarization controller. LD: laser diode. WDM: wavelength division multiplexer. EDF: Erbium doped fiber. ISO: isolator. The inset shows the transferred BP on the optical fiber connector. The scale bar: 60 μm. The red point in the center indicates the 10μm-diameter optical fiber core, with the outside concentric grey circle of the 125-μm fiber cladding. The area marked by the pink dotted line specifies the transferred BP. (b) Output spectrum. (c) Pulse width and repetition rate as a function of pump power. (d) Output pulse train.

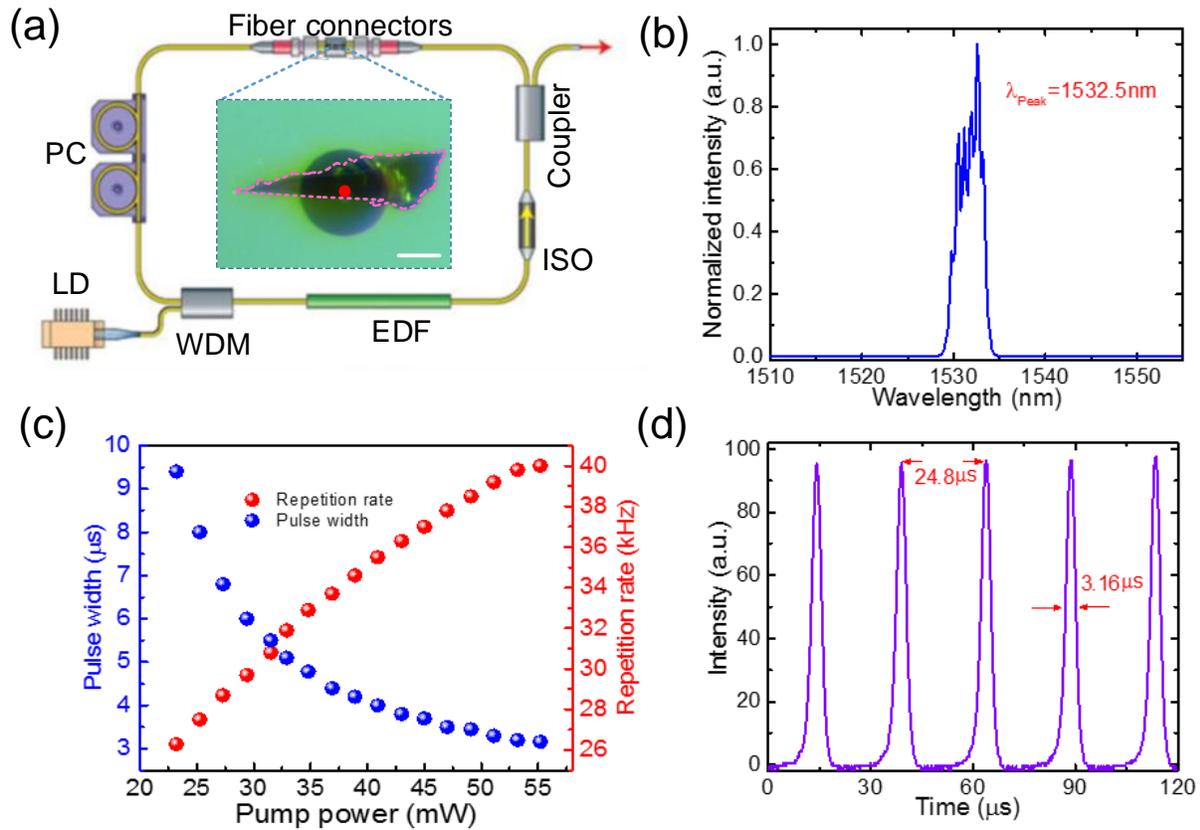

**Figure 4.** BP mode-locked fiber laser results: (a) Output spectrum. (b) Output autocorrelation trace, giving a pulse duration of ~786 fs. (c) Output pulse train. (d) Radio-frequency spectrum at the cavity fundamental repetition rate $f_0$ ($f_0$=14.7 MHz). The resolution bandwidth is 100 Hz.

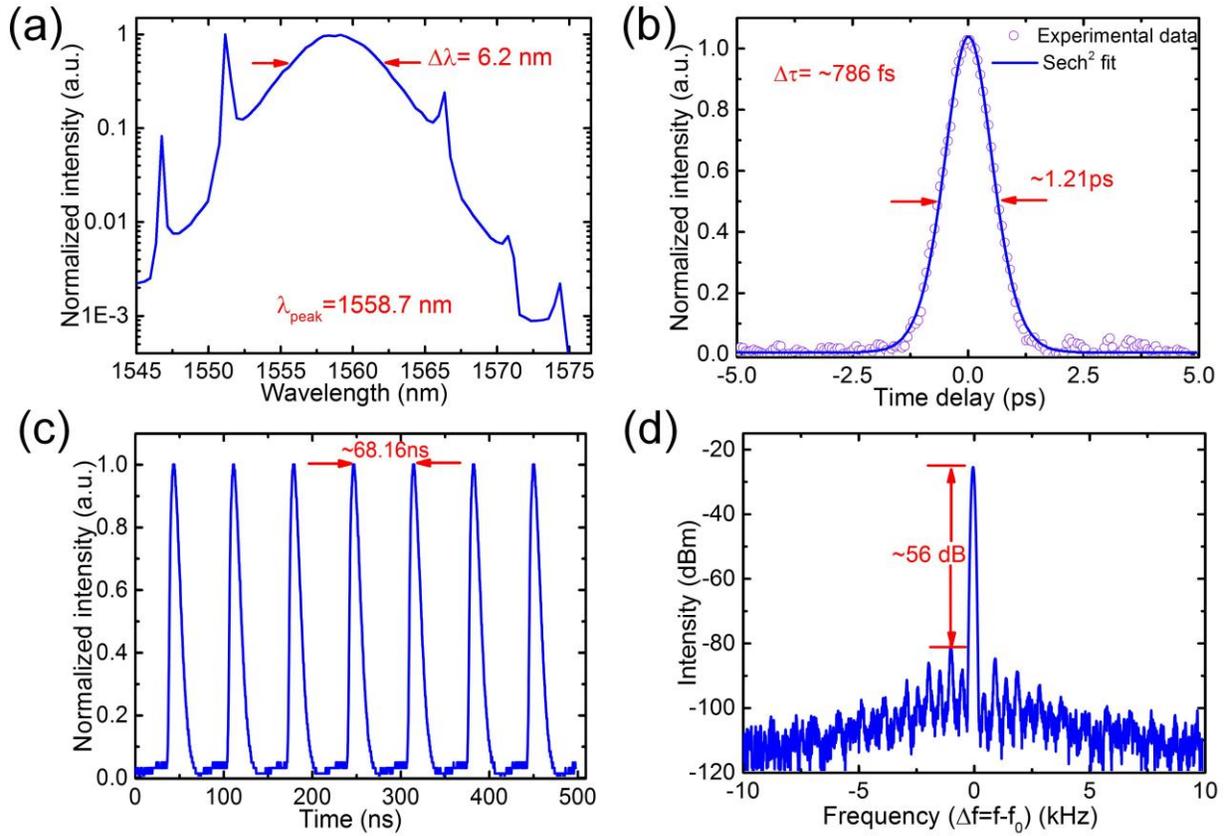

## AUTHOR INFORMATION

**Zhipei Sun**

Department of Micro- and Nanosciences, Aalto University, Tietotie 3, FI-02150 Espoo, Finland

Email: zhipei.sun@aalto.fi

**Note**

During the preparation of this manuscript, we became aware of two experimental works studying pulsed fiber lasers with BP on arXiv.org (arXiv: 1504.04731, arXiv: 504.07341).

**Funding Sources**

The authors acknowledge funding from Teknologiateollisuus TT-100, Academy of Finland (Grants: 276376, 284548, 285972), the European Union's Seventh Framework Programme (REA grant agreement No. 631610), China Scholarship Council, TEKES (NP-Nano), Jenny and Antti Wihuri Foundation, and Aalto University (Finland). H.J. thanks the two-week hostage of Cambridge Graphene Center, where he learnt the micro-positioning of micromechanically exfoliated layered materials. The authors also thank the provision of facilities and technical support by Aalto University at Micronova Nanofabrication Centre.